\documentclass[a4paper,UKenglish,cleveref,autoref,pdfa]{lipics-v2021}

\nolinenumbers%
\pdfoutput=1 %
\hideLIPIcs%

\title{Complexity Analysis for Call-by-Value Higher-Order Rewriting} %

\author{Cynthia Kop}
{Institute for Computing and Information Sciences, Radboud University, Nijmegen, The Netherlands \and \url{https://www.cs.ru.nl/~cynthiakop/index_en.html}}
{c.kop@cs.ru.nl}
{https://orcid.org/0000-0002-6337-2544}
{}

\author{Deivid Vale}
{Institute for Computing and Information Sciences, Radboud University, Nijmegen, The Netherlands \and \url{https://deividrvale.github.io}}
{D.Vale@cs.ru.nl}
{https://orcid.org/0000-0003-1350-3478}
{}

\authorrunning{C. Kop and D. Vale}

\Copyright{Cynthia Kop and Deivid Vale} %



\funding{This work is supported by the NWO TOP project
``Implicit Complexity through Higher-Order Rewriting'',
NWO 612.001.803/7571
and the NWO VIDI project ``Constrained Higher-Order Rewriting and Program Equivalence'', NWO VI.Vidi.193.075.
}

\EventEditors{John Q. Open and Joan R. Access}
\EventNoEds{2}
\EventLongTitle{42nd Conference on Very Important Topics (CVIT 2016)}
\EventShortTitle{CVIT 2016}
\EventAcronym{CVIT}
\EventYear{2016}
\EventDate{December 24--27, 2016}
\EventLocation{Little Whinging, United Kingdom}
\EventLogo{}
\SeriesVolume{42}
\ArticleNo{23}

\usepackage{mathtools,amssymb,amsfonts,amscd,amsbsy,}
\usepackage{microtype}

\usepackage{etex} %
\usepackage{tikz}
\definecolor{shade}{HTML}{F4F4FF}%
\newcommand\shademath[1]{
    \raisebox{-6pt}{
        \begin{tikzpicture}
            \node[fill=shade,rounded corners=3pt]{\(#1\)};
        \end{tikzpicture}
    }
}

\usepackage{bussproofs}

\usepackage{xcolor-material}
\usepackage{stmaryrd}

\usepackage{xparse}
\let\oldtheorem\theorem%
\RenewDocumentCommand{\theorem}{o}{%
  \IfNoValueTF{#1}
    {\oldtheorem}
    {\oldtheorem[#1]}%
  \normalfont{}
}

\let\olddefinition\definition%
\RenewDocumentCommand{\definition}{o}{%
  \IfNoValueTF{#1}
    {\olddefinition}
    {\olddefinition[#1]}%
  \normalfont{}
}

\let\oldlemma\lemma%
\RenewDocumentCommand{\lemma}{o}{%
  \IfNoValueTF{#1}
    {\oldlemma}
    {\oldlemma[#1]}%
  \normalfont{}
}

\usepackage{bm}

\newcommand{\sortFont}[1]{\mathsf{#1}}

\newcommand{\compFont}[1]{\mathsf{#1}}

\newcommand{\consFont}[1]{\mathsf{#1}}
\newcommand{\defFont}[1]{\mathsf{#1}}

\newcommand{\metaFont}[1]{\mathtt{#1}}

\newcommand{\nat}{\sortFont{nat}}
\newcommand{\lst}{\sortFont{list}}

\newcommand{\cost}{\compFont{c}}
\newcommand{\size}{\compFont{s}}
\newcommand{\leng}{\compFont{l}}
\newcommand{\mmax}{\compFont{m}}

\newcommand{\zero}{\consFont{0}}
\newcommand{\nil}{\consFont{nil}}
\newcommand{\suc}{\consFont{s}}
\newcommand{\cons}{\consFont{cons}}

\newcommand{\add}{\defFont{add}}
\newcommand{\double}{\defFont{dbl}}
\newcommand{\mult}{\defFont{mult}}

\newcommand{\map}{\defFont{map}}

\newcommand{\listvar}{q}

\newcommand{\asort}{\iota}

\newcommand{\atype}{\sigma}
\newcommand{\btype}{\tau}

\newcommand{\afun}{\defFont{f}}

\newcommand{\adata}{\consFont{c}}

\newcommand{\avar}{x}
\newcommand{\bvar}{y}

\newcommand{\aListVar}{xs}

\newcommand{\aFuncVar}{F}

\newcommand{\aterm}{s}
\newcommand{\bterm}{t}

\newcommand{\avalue}{v}

\newcommand{\num}[1]{\mathsf{#1}}

\newcommand{\aDom}{\mathcal{F}}

\newcommand{\costInt}[1]{\mathcal{C}_{#1}}
\newcommand{\costFInt}[1]{\mathcal{F}^\cost_{#1}}
\newcommand{\sizeInt}[1]{\mathcal{S}_{#1}}

\newcommand{\syntaxSig}{\mathbb{F}}
\newcommand{\signature}{\Sigma}

\newcommand{\dfdS}{\signature^{\mathtt{def}}}
\newcommand{\ctrS}{\signature^{\mathtt{con}}}
\newcommand{\var}{\mathbb{X}}
\newcommand{\terms}{\mathsf{T}(\syntaxSig,\var)}

\newcommand{\fvars}[1]{\mathtt{fv}(#1)}

\newcommand{\sortset}{\mathbb{B}}
\newcommand{\simpletypeset}{\mathbb{T}(\sortset)}

\newcommand{\rules}{\mathbb{R}}
\newcommand{\Nat}{\mathbb{N}}

\newcommand{\rulesArrow}{\to}

\newcommand{\arrz}{\rulesArrow}

\newcommand{\arrv}{\to_v}

\newcommand{\interpret}[1]{{\llbracket{} #1 \rrbracket}}
\newcommand{\ainterpret}[1]{\interpret{#1}_\alpha^{\funcinterpret{}}}
\newcommand{\typeinterpret}[1]{{\llparenthesis{} {#1} \rrparenthesis}}

\newcommand{\typecount}[1]{K(#1)}

\newcommand{\funcinterpret}[1]{\mathcal{J}_{#1}}

\newcommand{\costGt}{>}

\newcommand{\pair}[1]{\left\langle #1 \right\rangle}

\newcommand{\defText}[1]{\textbf{#1}}

\newcommand{\tuple}[1]{\bm{#1}}
\NewDocumentCommand{\typeVec}{m o}{
    \IfValueTF{#2}{
        {}^{#2}\bm{#1}
    }{
        \bm{#1}
    }
}

\newcommand{\fatlambda}{\lambda\!\!\!\lambda}

\newcommand{\lamT}[2]{\lambda{} {#1} {.\,} {#2}}

\newcommand{\app}{\,}
\newcommand{\semApp}{\cdot}

\newcommand{\unitSet}{\metaFont{unit}}
\newcommand{\unit}{\mathtt{u}}

\newcommand{\hasType}{\mathbin{:}}

\NewDocumentCommand{\irc}{o}{
    \IfValueTF{#1}{
        \metaFont{irc}_{\rules_{#1}}
    }{
        \metaFont{irc}_{\rules}
    }
}

\newcommand{\asub}{\gamma}

\newcommand{\dht}[1]{\metaFont{dh}_\rules(#1)}
\newcommand{\ar}{\metaFont{ar}}

\newcommand{\arrtype}{\Rightarrow}
\newcommand{\arrfunc}{\longrightarrow}
\newcommand{\arrfuncwm}{\Longrightarrow}
\NewDocumentCommand{\intKey}{o}{%
    \IfValueTF{#1}{
        \mathcal{J}_{\sortset/#1}
    }{
        \mathcal{J}_{\sortset}
    }
}

\newcommand{\cs}{cost--size} %
\newcommand{\cstitle}{Cost--Size} %

\begin{document}

\maketitle

\begin{abstract}
    In this short paper,
    we consider a form of higher-order rewriting with a call-by-value evaluation
    strategy so as to model call-by-value programs.
    We briefly present a \cs{} semantics to call-by-value rewriting:
    a class of algebraic interpretations that map terms to tuples
    that bound both the reductions' cost and the size of normal forms.
\end{abstract}

\section{Introduction}\label{sec:intro}

This short paper is a brief exposition of the conference paper
``\cstitle{} Semantics for Call-by-Value Higher-Order Rewriting''
recently published at FSCD 2023~\cite{kop:vale:23}.
We study \emph{complexity} in this work, which in the context of term rewriting
is typically understood as the number of steps needed to reach a normal from when
starting in terms of a certain shape and size.
A natural way to determine these bounds is by adapting techniques for proving
termination to deduce the complexity.
There is a myriad of works following this idea.
To mention a few,
see~\cite{
    baillot:lago:16,
    bonfante:cichon:marion:touzet:01,
    cichon:lescanne:92,
    hof:01,
    hof:lau:89}
for interpretation methods,~\cite{bonfante:et-al:01,hof:92,weiermann:95}
for lexicographic and path orders,
and~\cite{nao:moser:08,noschinski:emmes:jurgen:11} for dependency pairs.
However, those ideas are focused on \emph{first-order} term rewriting.
The literature on the complexity of \emph{higher-order} rewriting is scarce.
While there is a lot of work studying the complexity of functional
programs~\cite{ava:lag:17,danner:et-al:15,kahn:hoffman:20,niu:hoffman},
this work uses quite different ideas from the methods developed for term rewriting.
It would be beneficial to combine these ideas.

In a previous work~\cite{kop:vale:21},
we introduced an extension of the method of
\emph{weakly monotonic algebras}~\cite{fuh:kop:12,pol:96} to \emph{tuple
interpretations}.
This work deals with complexity analysis of higher-order rewriting
in the context of full rewriting, i.e.,
no choice for the evaluation strategy.
The idea of algebraic interpretations is to choose an interpretation domain
\(A\), and interpret terms \(s\) as elements \(\interpret{s}\) of \(A\)
compositionally in such a way that whenever \(\aterm \arrz \bterm \) we have
\(\interpret{\aterm} \costGt \interpret{\bterm}\).
Hence, a rewriting step on terms implies a strict decrease on \( A \).
The defining characteristic of tuple interpretations is to split
the complexity measure into abstract notions of cost and size.
This coincides with ideas often used in resource analysis of functional
programs~\cite{ava:lag:17,danner:et-al:15}.
This is a popular idea, as a very similar approach was
introduced for first-order rewriting around the same time~\cite{yamada:21}.

\section{Preliminaries}\label{sec:prel}

The formalism we consider here is a style of simply typed lambda
calculus extended with function symbols and rules.
The matching mechanism is modulo alpha,
and beta reduction is included in the rewriting relation.

Let \( \sortset \) be a nonempty set of \emph{base types}.
The set \( \simpletypeset \) of
\emph{simple types} over \( \sortset \)
is generated by the grammar:
\({ \simpletypeset
\coloneq
\sortset \mid \simpletypeset \arrtype \simpletypeset
}\).
As usual, we assume that the \( \arrtype \) type constructor is right-associative.
A \emph{signature} \( \syntaxSig \) is a triple
\( {(\sortset, \signature, \ar)} \) where
\( \sortset \) is a set of base types,
\( \signature \) is a nonempty finite set of symbols,
and \( \ar \) is a function \( \ar : \signature \arrfunc \simpletypeset \).
We postulate, for each type \( \atype \),
the existence of a nonempty set \( \var_\atype \)
of countably many variables.
Furthermore,
we impose that
\( \var_\atype \cap \var_\btype = \emptyset \)
whenever \( \atype \neq \btype \) and
let \( \var \) denote the family of sets
\( {(\var_\atype)}_{\atype \in \simpletypeset} \)
indexed by \( \simpletypeset \)
and assume that \( \signature \cap \var = \emptyset \).

The set \( \terms \) ---
of terms built from \( \syntaxSig \) and \( \var \) ---
collects those expressions \( \aterm \)
for which the judgment \( \aterm \hasType \atype \)
can be deduced using the following rules:
\begin{prooftree}
    \AxiomC{\( \avar \in \var_\atype \)}
    \UnaryInfC{\( \avar \hasType \atype \)}
    \DisplayProof{}
    \hfill
    \AxiomC{\( \afun \in \signature \)}
    \AxiomC{\( \ar(\afun) = \atype \)}
    \BinaryInfC{\( \afun \hasType \atype \)}
    \DisplayProof{}
    \hfill
    \AxiomC{\( \aterm \hasType \atype \arrtype \btype \vphantom{\var_\atype}\)}
    \AxiomC{\( \bterm \hasType \atype \)}
    \BinaryInfC{\((\aterm \app \bterm ) \hasType \btype \)}
    \DisplayProof{}
    \hfill
    \AxiomC{\( \avar \in \var_\atype \)}
    \AxiomC{\( \aterm \hasType \btype \vphantom{\var_\atype}\)}
    \BinaryInfC{\( {(\lamT{\avar}{\aterm})} \hasType \atype \arrtype \btype \)}
\end{prooftree}

We assume the usual \( \lambda \)-calculus association and precedence scheme for
application and abstraction.
We shall remove unnecessary parentheses and write terms following those rules.
Application of substitutions is defined as expected.

\subparagraph*{Call-by-Value Higher-order Rewriting}
A \emph{rewrite rule} \( \ell \to r \) is a pair of terms of the same type
such that \( \ell = \afun \app \ell_1 \dots \ell_k \) and
\( \fvars{r} \subseteq \fvars{\ell} \),
here \( \fvars{\aterm} \) is the function mapping a term to the set of its free
variables.
A \emph{term rewriting system} (TRS)
\( \rules \) is a set of rules.
In this paper, we are interested in a restricted evaluation strategy, which
limits reduction to terms whose immediate subterms are \emph{values}:

\begin{definition}\label{def:value}
A term \( \aterm \) is a \textit{value} whenever \( \aterm \) is:
\begin{itemize}
    \item of the form \( \afun \app \avalue_1 \ldots \avalue_n \),
    with each \( v_i \) a value and
    there is no rule \( \afun \app \ell_1 \ldots \ell_k \arrz r \) %
    with \( k \leq n \);
    \item an abstraction, i.e., \( \aterm = \lamT{\avar}{\bterm} \).
\end{itemize}
\end{definition}

Every rewrite rule
\( \ell \rulesArrow r \)
\emph{defines} a symbol \( \afun \), namely,
the head symbol of \( \ell \).
For each \( \afun \in \signature \),
let \( \rules_\afun \) denote the set of rewrite rules
that define \( \afun \) in \( \rules \).
A symbol \( \afun \in \signature \) is a
\emph{defined symbol} if
\( \rules_\afun \neq \emptyset \).
A \emph{constructor symbol} is a symbol \( \adata \in \signature \)
such that %
\( \rules_\adata = \emptyset \).
We let \( \dfdS \) be the set of defined symbols
and \( \ctrS \) the set of constructor symbols.
Hence, \( \signature = \dfdS \uplus \ctrS \).
A \emph{ground constructor term} is a term \(\adata \app \aterm_1 \ldots
\aterm_n \) with \(n \geq 0\), where each \( \aterm_i \) is a ground constructor term.

Notice that by definition ground constructor terms are values since there is no
rule $\adata\ \ell_1\ \dots\ \ell_k \arrz r$ for any $k$ if $\adata \in \ctrS$.
More complex values include partially applied functions and lambda-terms; for
example, \( \add \app \zero \) or a list of functions \( [\add \app \zero;
\lambda x. x ; \mult \app \zero; \double] \).

\begin{definition}\label{def:ho-cbv-rw}
The \defText{higher-order weak call-by-value rewrite relation} \( \arrv \)
induced by \( \rules \) is defined as follows:
\begin{itemize}
    \item \( \afun \app (\ell_1 \asub) \dots (\ell_k \asub) \arrv r\asub \),
        if \( \afun \app \ell_1 \dots \ell_k \arrz r \in \rules \) and each
        \( \ell_i \asub \) is a value;
    \item \( (\lamT{\avar}{\aterm}) \app \avalue \arrv \aterm [\avar \coloneq \avalue] \),
        if \( \avalue \) is a value;
    \item \( \aterm \app \bterm \arrv \aterm' \app \bterm \)
        if \( \aterm \arrv \aterm' \); and
        \( \aterm \app \bterm \arrv \aterm \app \bterm' \)
        if \( \bterm \arrv \bterm' \).
\end{itemize}
\end{definition}

\begin{example}\label{ex:ho-toy-trs}
    Let us consider two simple examples of functions encoded as rules.
    The first is \( \map \), which applies a function
    \( \aFuncVar \hasType \nat \arrtype \nat \).
    \begin{align*}
        & \map \app \aFuncVar \app \nil \arrz \nil & &
        \add \app \avar \app \zero \arrz \zero
        \\
        & \map \app \aFuncVar \app (\cons \app \avar \app \aListVar) \arrz
        \cons \app (\aFuncVar \app \avar) \app
        (\map \app \aFuncVar \app \aListVar) & &
        \add \app \avar \app (\suc \app \bvar) \arrz \suc \app (\add \app \avar \app \bvar)
    \end{align*}
\end{example}

\newpage
\section{\cstitle{} Semantics for Types and Terms}\label{sec:ituple}

An interpretation of types is a function \( \typeinterpret{\cdot} \)
that maps each type \( \atype \in \simpletypeset \)
to a well-founded set \( \typeinterpret{\atype} \),
the \cs{} interpretation of \( \atype \).
In order to define such a function,
we need an \textit{interpretation key} function
\( K : \sortset \arrfunc \Nat \) mapping
base types \( \asort \) to a number \( \typecount{\asort} \).
This number sets the ``length'' of the tuples
in the size interpretation of \( \asort \).

\begin{definition}[Interpretation of Types]\label{def:int-ty}
    We define for each type \( \atype \)
    the \defText{\cs{} tuple interpretation} of \( \atype \)
    as the set \( \typeinterpret{\atype} = \costInt{\atype} \times \sizeInt{\atype} \)
    where \( \costInt{\atype} \) and \( \sizeInt{\atype} \)
    are defined as follows:
    \begin{align*}
        \costInt{\atype}
        &=
        \Nat \times \costFInt{\atype}
        &
        \sizeInt{\asort}
        & = \Nat^{\typecount{\asort}} \\
        \costFInt{\asort} & = \unitSet                                                                &
        \sizeInt{\atype \arrtype \btype}  & = \sizeInt{\atype} \arrfuncwm \sizeInt{\btype}                                                              \\
        \costFInt{\atype \arrtype \btype} & = (\costFInt{\atype} \times \sizeInt{\atype}) \arrfuncwm \costInt{\btype}
    \end{align*}
\end{definition}
The set \( \typeinterpret{\atype} \) is ordered component-wise.
With that this interpretation of types is well-founded,
which was proved in the full version of this paper.
Next,
we need an \textit{application operator} for applying \cs{} tuples.
More precisely, given a type \( \atype \arrtype \btype \)
and \cs{} tuples
\( \tuple{f} \in \typeinterpret{\atype \arrtype \btype} \) and
\( \tuple{x} \in \typeinterpret{\atype} \), we define
the application of \( \tuple{f} \) to \( \tuple{x} \) as follows.

\begin{definition}\label{def:semantic-app}
    Let \( \atype \arrtype \btype \) be an arrow type,
    \( \tuple{f} =
    \pair{(n, f^\cost),f^\size} \in \typeinterpret{\atype \arrtype \btype} \),
    and
    \( \tuple{x} = \pair{(m, x^\cost), x^\size} \in \typeinterpret{\atype} \).
    The \defText{semantic application} of \( \tuple{f} \) to \( \tuple{x} \),
    denoted \( \tuple{f} \semApp \tuple{x} \), is defined by:
    \begin{center}
        let \( f^\cost(x^\cost, x^\size) = (k,h) \); then
        \( \pair{(n, f^\cost), f^\size} \semApp \pair{(m, x^\cost), x^\size} =
        \pair{(n + m + k, h), f^\size(x^\size)}
        \)
    \end{center}
\end{definition}

An interpretation of a signature
\( \syntaxSig = {(\sortset, \signature, \ar)} \)
interprets the base types in \( \sortset \)
and each \( \afun \in \signature \) of arity \( \ar(\afun) = \atype \)
as an element of \( \typeinterpret{\atype} \)
which is constructed by Definition~\ref{def:int-ty}.

\begin{definition}\label{def:cs-tuple-int}
    A \defText{\cs{} tuple interpretation} \( \aDom \) for a signature
    \( \syntaxSig = {(\sortset, \signature, \ar)} \)
    consists of a pair of functions \( (K, \funcinterpret{\signature}) \) where
    \begin{itemize}
        \item \( K \) is a type interpretation key,
            which maps each base type \( \asort \) to its tuple dimension \( \typecount{\asort} \)
        \item \( \funcinterpret{\signature} \) is an
        \emph{interpretation of symbols} in \( \signature \)
        which maps each \( \afun \in \signature \)
        with \( \ar(\afun) = \atype \)
        to a \cs{} tuple in
        \( %
        \typeinterpret{\atype} \),
        where \( \typeinterpret{\atype} \)
        is built using \( K \) in
        Definition~\ref{def:int-ty}.
    \end{itemize}
\end{definition}
In what follows we slightly abuse notation by writing \( \funcinterpret{\afun} \)
for \( \funcinterpret{\signature}(\afun) \)
and just \( \funcinterpret{} \) for \( \funcinterpret{\signature} \).

\begin{example}\label{ex:int-constructors}
As a first example of interpretation,
let us interpret the data constructors from \cref{ex:ho-toy-trs}.
Recall that \( \zero \hasType \nat, \suc \hasType \nat \arrtype \nat \)
are the constructors for \( \nat \).
We then set \( \typecount{\nat} = 1 \).
\begin{align*}
    \funcinterpret{\zero} &=
    \pair{\shademath{(0,\unit)}, 1}
    &
    \funcinterpret{\suc} &=
    \pair{\shademath{(0, \fatlambda x. (0, \unit))}, \fatlambda x. x + 1}
    \phantom{00000000000000}
\end{align*}
The highlighted cost components for the constructors are filled with zeroes.
That is because in the rewriting cost model data values do not
fire rewriting sequences.
Intuitively, the \emph{cost number} for $\zero$ is 0,
(because it is a value),
the \emph{cost function} is the unit element \( \unit \in \unitSet \),
(because it has base type),
and \emph{size component} is $1$ (since we chose a notion of size for terms
of type $\nat$ to mean ``number of symbols'').
The cost number for $\suc$ is $0$, the cost function is the constant function mapping
to $0$, and the size component is the function \( \fatlambda x. x + 1 \)
in \( \sizeInt{\nat \arrtype \nat} \).
We interpret the constructors for \( \lst \), i.e., \( \nil \) and \( \cons \),
following the same principle, with \( K(\lst) = 2 \).
We write a size tuple \( \listvar \) in \( \sizeInt{\lst} \) as
\( (\listvar_\leng, \listvar_\mmax) \) since
the first component is to mean the length of the list and the second a bound
on the size of its elements.
\begin{align*}
     \funcinterpret{\nil} &=
     \pair{\shademath{(0,\unit)}, (0,0)}
     &
     \funcinterpret{\cons} &=
     \pair{
     \shademath{(0, \fatlambda x. (0, \fatlambda \listvar. (0, \unit)))},
     \fatlambda x \listvar. (q_\leng + 1, \max(x, q_\mmax))
     }
\end{align*}
The highlighted cost components are filled with zeroes for lists as well.
Size components are interpreted following the semantics we set for
the two size components lenght and maximum element size, respectively.
\end{example}

The next step is to extend the interpretation of a signature
\( \syntaxSig \) to the set of terms.
But first, we define \emph{valuation functions} to interpret the variables
in \( x \hasType \atype \)
as elements of \( \typeinterpret{\atype} \).

\begin{definition}\label{def:valuation}
    A \defText{\cs{} valuation} \( \alpha \) is a function that maps each
    \( \avar \hasType \atype \) to a cost-size tuple
    in \( \typeinterpret{\atype} \) such that:
    \begin{itemize}
        \item \( \alpha(x) = \pair{(0, \unit), x^\size} \),
            for all \( \avar \in \var \) of base type;
        and  \( \alpha(F) = \pair{(0, F^\cost), F^\size} \)
            when \( F :: \atype \arrtype \btype \).
    \end{itemize}
\end{definition}

\begin{definition}\label{def:int-tm}
    Assume given a signature \( \syntaxSig = (\sortset, \signature, \ar) \) and
    its \cs{} tuple interpretation \( \aDom = (K, \funcinterpret{} ) \)
    together with a valuation \( \alpha \).
    The \defText{term interpretation} \( \ainterpret{\aterm} \) of \( \aterm \)
    under \( \funcinterpret{} \) and \( \alpha \)
    is defined by induction on the structure of \( \aterm \) as follows:
    \[
    \begin{array}{rclcrclcrcl}
        \ainterpret{\avar} & = & \alpha(\avar) & \quad &
        \ainterpret{\afun} & = & \funcinterpret{\afun} & \quad &
        \ainterpret{\aterm \app \bterm} & = & \ainterpret{\aterm} \semApp \ainterpret{\bterm} \\
        \multicolumn{11}{l}{
          \ainterpret{\lamT{\avar}{\aterm}}
        = \pair{
            \left(
            0,
            \fatlambda d. (1 +
            \pi_{11}(\interpret{\aterm}^{\funcinterpret{}}_{[x \coloneq d]\alpha}),
            \pi_{12}(\interpret{\aterm}^{\funcinterpret{}}_{[x \coloneq d]\alpha}))
            \right),
            \fatlambda d^\size.
            \pi_2(\interpret{\aterm}^{\funcinterpret{}}_{[x \coloneq (\underline{0},d)]\alpha})
        },
        }
    \end{array}
    \]
    where
    \( \pi_i \) is the projection on the ith-component
    and \( \pi_{ij} \) is the composition \( \pi_j \circ \pi_i \),
    and $\underline{0}$ is a cost function of the form $\fatlambda x_1.(0,\fatlambda x_2
    \dots (0,\unit)\dots)$.
    If $d = (d^c,d^s)$, the notation \( [x \coloneq d]\alpha \) denotes the valuation that
    maps $x$ to $\pair{(0,d^c),d^s}$ and every other variable $y$ to $\alpha(y)$.
\end{definition}
We write \( \interpret{\aterm} \) for \( \ainterpret{\aterm} \)
whenever \( \alpha \) and \( \funcinterpret{} \) are universally quantified
or clear from the context.

The interpretation for abstractions may seem baroque, but can be understood as follows:
an abstraction is a value, so its cost number is $0$.  The cost of applying that
abstraction on a value $v$ is $1$ plus the cost number for $s[x:=v]$ -- which is
obtained by evaluating $\interpret{\aterm}^{\funcinterpret{}}_{[x \coloneq d]\alpha}$
if $d$ is the cost function/size pair for $v$.  The cost \emph{function} of this
application is exactly the cost function of $s[x:=v]$.
The \emph{size} of an abstraction $\lambda x.s$ is exactly the function that takes a
size and maps it to the size interpretation of $s$ where $x$ is mapped to that size.
Technically, to obtain the size component of
$\interpret{\aterm}^{\funcinterpret{}}_{[x\coloneq d]\alpha}$ we also need a cost component, but by definition, this component
does not play a role, so we can safely choose an arbitrary pair $\underline{0}$ in
the right set.

\begin{example}\label{ex:int-data-tm}
We continue with \cref{ex:int-constructors} by interpreting ground constructor
terms fully.
A ground constructor term \( d \) of type \( \nat \) is of the form
\( \suc \app (\suc \dots (\suc \app \zero)\dots) \) where the number \( n \in \Nat \)
is represented by \( n \) successive applications of \( \suc \) to \( \zero \).
Let us write \( \num{n} \) as shorthand notation for such terms.
Similarly, for ground constructor terms of type \( \lst \),
we write \( [\num{n}_1; \dots; \num{n}_k ] \) for the term
\( \cons \app \num{n}_1 \app \dots (\cons \app \num{n}_k \app \nil) \).
The empty list constructor \( \nil \) is written as \( [] \) in this notation.
Hence, the \cs{} interpretation of \( \num{3} \hasType \nat \) is given by:
\[
    \interpret{\num{3}} =
    \interpret{\suc \app (\suc \app (\suc \app \zero))} =
    \interpret{\suc} \semApp
    (\interpret{\suc} \semApp
    (\interpret{\suc} \semApp \interpret{\zero})) =
    \pair{\shademath{(0,\unit)}, 4}.
\]
Consider, for instance, the list \( [\num{1}; \num{7}; \num{9}] \).
Its \cs{} interpretation is given by:
\[ \interpret{[1; 7; 9]} =
    \interpret{
        \cons \app \num{1} \app
        (\cons \app \num{7} \app
        (\cons \app \num{9} \app \nil))
    } \phantom{0} =
    \pair{\shademath{(0, \unit)}, (3, 10)}.
\]
The important information we can extract from such interpretations is their size
component.
Indeed,
\( \interpret{3}^\size = 4 \) counts the number of constructor symbols in the
term representation \( \num{3} \)
and
\( \interpret{[1; 7; 9]}^\size = (3, 10) \)
gives us the length and an upper bound on the size of each element in
\( [1; 7; 9] \).
The size interpretation for the constructors of \( \nat \) and \( \lst \)
correctly capture our notion of ``size'' given earlier.
\end{example}

We give a concrete \cs{} interpretation for \( \map \) and \( \add \) below:
\[
    \funcinterpret{\add} =
    \pair{
        \shademath{(0, \fatlambda x. (0, \fatlambda y. (y^\size, \unit)))},
        \fatlambda x y. x + y
    }.
\]
\[
    \funcinterpret{\map} = \pair{
        \shademath{
            (0, \fatlambda F. (0, \fatlambda q.
            (q_\leng + F^\cost(\unit, q_\mmax)q_\leng + 1, \unit)))
        },
        \fatlambda F q. (q_\leng, F(q_\mmax))
    },
\]

\section{Complexity Analysis of Call-by-Value Rewriting}\label{sec:comp-analysis}

Since our analysis is quantitative,
our goal is not merely to find tuple interpretations that prove termination
but also ones that provide ``good'' upper bounds on the complexity of reducing
terms.
To start, we will extend the notion of \emph{derivation height} to our setting:

\begin{definition}
The weak call-by-value \defText{derivation height} of a term $\aterm$, notation
$\dht{\aterm}$, is the largest number $n$ such that $\aterm \arrv \aterm_1
\arrv \dots \arrv \aterm_n$.
\end{definition}

This notion is defined for all terms when the TRS is terminating.
The methodology of weakly monotonic algebras offers a systematic way to derive
bounds for the derivation height of a given term:

\begin{lemma}\label{lem:interpretdh}
If $\interpret{\aterm} = \pair{(n,F^\cost),F^\size}$, then $\dht{\aterm} \leq n$.
\end{lemma}

As an illustration of how this is used, let us
complete the interpretation of
\cref{ex:ho-toy-trs}.
We start with the system \( \rules_\add \).
We will use the type and constructor interpretations as given
in \cref{ex:int-constructors}.
The rules in \( \rules_\add \)
suggest the following \cs{} interpretation:
\[
    \funcinterpret{\add} =
    \pair{
        \shademath{(0, \fatlambda x. (0, \fatlambda y. (y^\size, \unit)))},
        \fatlambda x y. x + y
    }.
\]

Notice that the (highlighted) cost component of \( \funcinterpret{\add} \)
suggest a linear cost measure for computing with \( \add \).
We also set the intermediate numeric components in the cost tuple to zero.
The reason for this choice is that in a cost tuple
\( \costInt{\atype} = \Nat \times \costFInt{\atype} \),
the numeric component \( \Nat \)
captures the cost of partially applying terms, which is $0$ in this case.

Now, consider the partially applied term
\( \aterm = \add \app ( \add \app \num{2} \app \num{3} ) \)
(of type \( \nat \arrtype \nat \)).
Intuitively, the cost of reducing this term to normal form, is the cost of
reducing the subterm \( \add \app \num{2} \app \num{3} \) to \( \num{5} \),
since the partially applied term \( \add \app \num{5} \) cannot be reduced.
Hence, \( \dht{\aterm} = 4 \).  This is also the bound we find through
interpretation:
\begin{align*}
    \interpret{\aterm}
    &=
    \interpret{\add} \semApp
    (
        \interpret{\add} \semApp
        \interpret{\num{2}}
        \semApp \interpret{\num{3}}
    )\\
    &=
    \interpret{\add} \semApp \pair{(4, \unit), 7}\\
    &=
    \pair{
        \shademath{(4, \fatlambda y. (y^\size, \unit))},
        \fatlambda y. 7 + y
    }.
\end{align*}
While in this case the upper bound we find is tight, this is not always the case; for
instance $\interpret{\add\ \num{0}\ (\add\ \num{0}\ \num{0})} = \pair{(3,\unit),
3}$, even though $\dht{\add\ \num{0}\ (\add\ \num{0}\ \num{0})} = 2$.
We could obtain a tight upper bound by choosing a different interpretation,
but this is also not always possible.

With this observation,
we get a framework that provides us with a systematic approach
to establish bounds to the complexity of weak call-by-value systems.
The difficulty now lies in developing techniques
to find suitable interpretation shapes.
For instance,
a first example of a higher-order function over lists is that of \( \map \).
We give a concrete \cs{} interpretation for \( \map \) below:
\[
    \funcinterpret{\map} = \pair{
        \shademath{
            (0, \fatlambda F. (0, \fatlambda q.
            (q_\leng + F^\cost(\unit, q_\mmax)q_\leng + 1, \unit)))
        },
        \fatlambda F q. (q_\leng, F(q_\mmax))
    },
\]
The highlighted cost component accounts for
\( q_\leng \) possible \( \beta \) steps,
the cost of applying the higher-order argument \( F \) over the list \( q \)
is bounded by \( F^\cost(\unit, q_\mmax) q_\leng \) since \( F^\cost \)
is assumed to be weakly monotonic,
and the unitary component is for dealing with the empty list case.

\section{Conclusions}\label{sec:conclusions}
In this short paper we briefly discussed an interpretation method for higher-order rewriting
with weak call-by-value reduction.
In this approach, we build on existing work defining tuple interpretations
\cite{kop:vale:21,yamada:21}, but restrict the evaluation strategy,
and define a \cs{} semantics for types and terms which generate a whole new class of \cs{} semantic techniques that can be used to reason about the complexity of weak call-by-value systems.

\bibliography{references}

\end{document}